%% file: main.tex
\documentclass[sigconf, 9pt, nonacm]{acmart}

\usepackage{balance}
\usepackage{hyperref}
\usepackage{url}
\usepackage{svg}
\usepackage{graphicx}
\usepackage{footnote}
\usepackage[flushleft]{threeparttable}

\newcommand{\scum}{SC\textmu M}

\newcommand{\plutosens}[1]{-82#1}
\newcommand{\bestsens}[1]{\plutosens{#1}}

\begin{document}

\author{Brandon P. Hippe*}
\author{David C. Burnett**}
\affiliation{
    \institution{Villanova University}
    \city{Villanova, PA} \country{USA}
}
\email{*bhippe@villanova.edu}
\email{**david.burnett@villanova.edu}

\author{Jacob N. Louie*}
\author{Tengfei Chang**}
\affiliation{
    \institution{Hong Kong University of Science and Technology, Guangzhou}
    \city{Guangzhou} \country{China}
}
\email{*jlouie475@connect.hkust-gz.edu.cn}
\email{**tengfeichang@hkust-gz.edu.cn}

\author{Dingyu Zhou*}
\affiliation{
    \institution{Portland State University}
    \city{Portland, OR} \country{USA}
}
\email{*ezhou@pdx.edu}

\author{Alfonso Cortés}
\author{Filip Maksimovic}
\affiliation{
    \institution{Inria, Paris}
    \city{Paris} \country{France}
}
\renewcommand{\shortauthors}{Hippe, et al.}

\title{A Digital Twin Platform Enabling Monolithic Crystal-Free Bluetooth Low Energy Single-Chip Sensor Motes}

\begin{abstract}
  Low-power wireless-capable systems-on-chips (SoCs) are critical for researching many of our current environmental issues. The scale at which these devices are needed for many applications necessitates innovation in their design to reduce the various capital and labor costs involved with operating an extensive sensor network. This can be difficult for devices with novel wireless architectures, as many emerging architectures lack commercially available development platforms. This makes pre-silicon validation challenging, and the impact of a failed tapeout is unacceptable when the cost is of primary concern for these devices. In this work, we propose a digital twin ecosystem for Bluetooth Low-Energy (BLE) with physical-layer (PHY) control intended for novel device development and demonstrated through use with crystal-free single-chip sensor motes. We present this system operating with multiple RF front ends and digital baseband implementations, including a commercially available Software Defined Radio (SDR) with synthesized RTL and embedded firmware, along with an existing crystal-free SoC front end and FPGA digital baseband. These configurations are shown to be capable of communicating sensor data with commercially available BLE devices and achieving receiver sensitivities up to \bestsens{~dBm}, exceeding the minimum BLE specification. This approach is extendable to other hardware and communication protocols and promises to enable inexpensive, reusable validation and verification tools for novel wireless devices.
\end{abstract}

\ccsdesc[500]{Hardware~Radio frequency and wireless circuits}
\ccsdesc[500]{Hardware~Wireless devices}
\ccsdesc[500]{Hardware~Sensor devices and platforms}
\ccsdesc[500]{Hardware~Digital signal processing}
\ccsdesc[500]{Hardware~Noise reduction}
\ccsdesc[500]{Hardware~Circuits power issues}
\ccsdesc[500]{Hardware~Analog and mixed-signal circuit optimization}
\ccsdesc[500]{Hardware~Hardware-software codesign}
\ccsdesc[500]{Hardware~Model checking}
\ccsdesc[500]{Hardware~Analog, mixed-signal and radio frequency test}
\ccsdesc[500]{Hardware~Emerging simulation}
\ccsdesc[500]{Hardware~Emerging tools and methodologies}
\ccsdesc[500]{Computer systems organization~System on a chip}

\keywords{Digital Twin, Bluetooth Low Energy, Crystal-Free, System-on-Chip, Simulation Platform}

\begin{teaserfigure}
    \centering
    \includegraphics[width=0.99\textwidth]{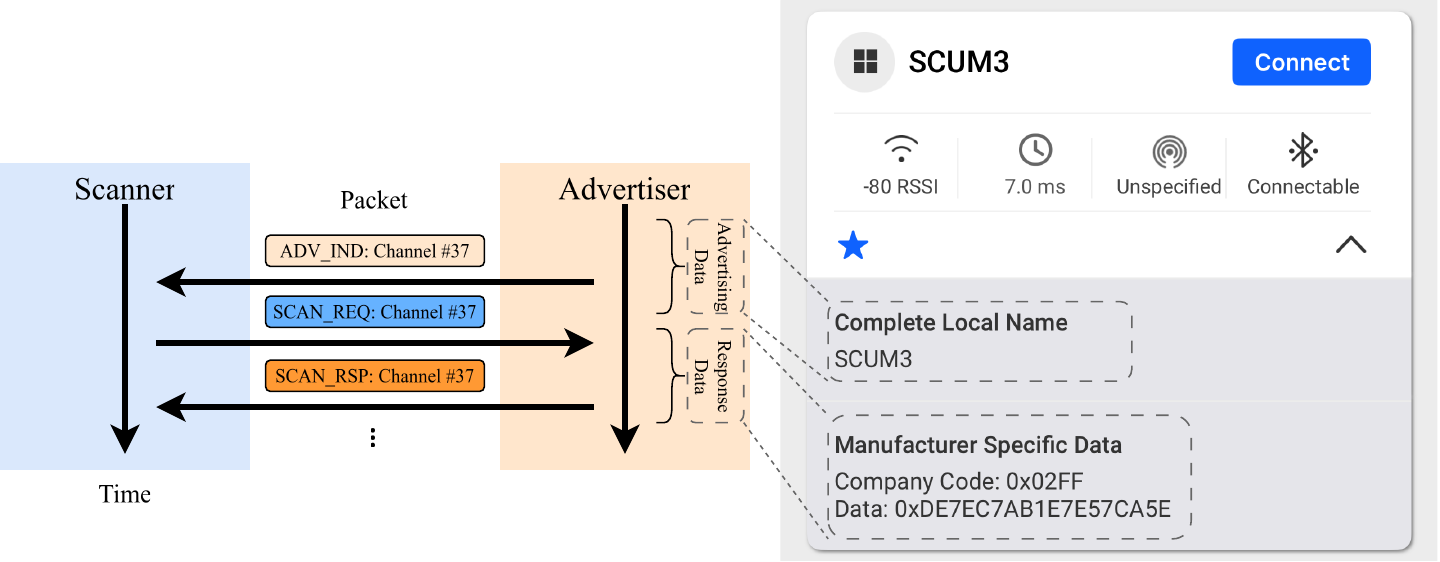}
    \caption{Proposed digital twin platform, with Crystal-Free \scum{} RF Front End and FPGA Digital Baseband, appearing as a connectable device in a BLE scanner app \cite{noauthor_simplicity_nodate}.}
    \label{fig:connectable}
\end{teaserfigure}

\maketitle

\section{Introduction} \label{Introduction}
\input{introduction}

\section{Design} \label{Design}
\input{design}

\section{Implementations \& Simulations} \label{Implementation}
\input{implementation}

\section{Demonstration \& Results} \label{Results}
\input{results}

\section{Conclusion} \label{Conclusion}
\input{conclusion}

\section{Acknowledgment}
The authors would like to thank the Berkeley Autonomous Microsystems Lab at the University of California, Berkeley, led by Prof. Kristofer S.J. Pister, for experimental hardware.

\bibliographystyle{ACM-Reference-Format}
\bibliography{references}

\end{document}

%% file: introduction.tex
\begin{figure*}[h!]
    \centering
    \includegraphics[width=0.90\textwidth]{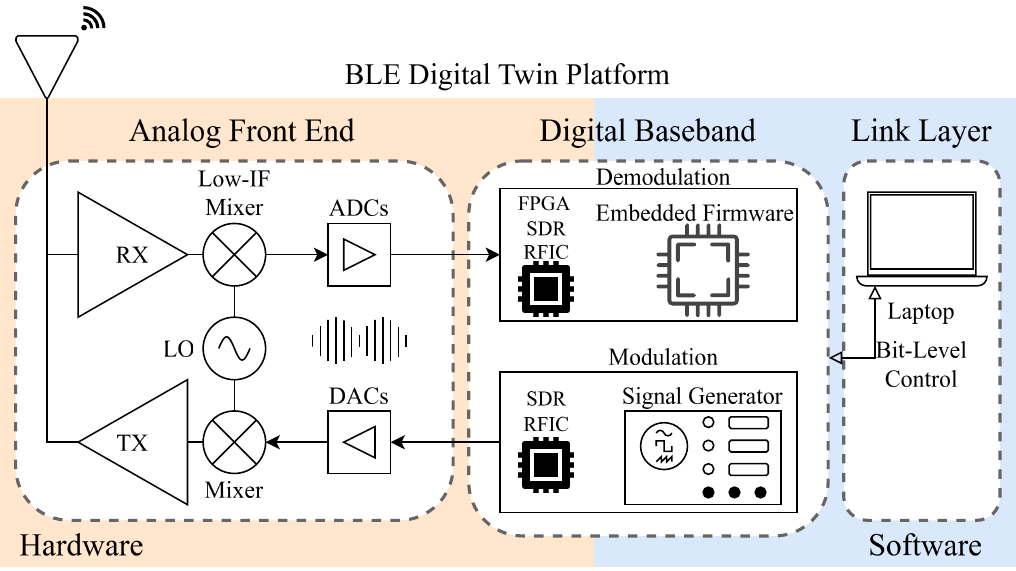}
    \caption{Block diagram of the proposed Bluetooth low-energy digital twin transceiver, showing how various subsystems can be implemented in multiple ways. The analog front end can be implemented using a commercially available software-defined radio or custom RF integrated circuit and supports traditional and crystal-free architectures. The digital baseband can be fulfilled in either software or hardware, allowing for easy design iterations and pre-silicon validation. A barebones link layer is implemented in software, allowing for bit-level control of the transceiver.}\label{fig:block_diagram}
\end{figure*}

The Internet of Things (IoT) promises improvements in everything from lifestyle to scientific research. 
However, there are still a variety of challenges before the Internet of Things becomes as ubiquitous as envisioned. 
Chief among these challenges is the cost of IoT devices. 
These costs come from various sources, including the labor required to deploy and maintain the device, the infrastructure required to operate the network, and the power consumed by the devices. 
Research is underway in novel system designs, energy reduction and harvesting techniques, and wireless communication strategies to help improve the adoption of the Internet of Things.

Our proposed system is a "Digital Twin" development ecosystem for Bluetooth Low-Energy (BLE) implementations using crystal-free wireless architectures, as is shown in Figure \ref{fig:block_diagram}. 
This system uses commercial-off-the-shelf (COTS) Software-Defined Radios (SDRs) for the base platform, with the ability to substitute any subsystem, such as new transmitter/receiver systems, analog front-ends, and digital baseband implementations, and gives this ecosystem wide-ranging compatibility with existing low-cost devices. 
This approach also allows subsystems of already fabricated devices to be used, even if they are incomplete or do not fully meet the specifications of the target design.
This streamlines the design process of new devices and allows for easier and more cost-efficient iteration during the design process. 
Although this work is intended for crystal-free BLE device development, this system is beneficial for other wireless architectures. Our proposed system is viable as a basic framework for low-level digital twins for other wireless protocols.

BLE is among the most widely used wireless communication protocols available, with billions of existing devices having BLE capabilities. 
BLE networks are minimally dependent on other infrastructure, so routers, network switches, cellular towers, repeaters, etc., are unnecessary.
BLE is also suitable for battery-powered devices or devices that utilize energy harvesting techniques. 
These advantages are crucial in minimizing infrastructure and maintenance costs of a large-scale IoT deployment. 

Although digital twins are common in the development of 5G and 6G wireless systems \cite{wu_digital_2021, nguyen_digital_2021, tang_survey_2022, khan_digital-twin-enabled_2022, lin_6g_2023, ahmadi_networked_2021, van_huynh_edge_2022, zhang_artificial_2022, bariah_digital_2023, pengnoo_digital_2020}, similar developments are lacking in the low-power IoT space. The proposed digital twin is targeted at improving the usability of crystal-free SoCs, such as the Single-Chip Micro Mote, or \scum{}. \scum{} \cite{maksimovic_crystal-free_2019, lovell_scm-v23_2024} is an attempt to design a monolithic SoC and aims to minimize the cost per device. 
Current iterations of \scum{} only require two external components (a power source and an antenna) to operate and feature a fully capable IEEE 802.15.4 transceiver, ARM Cortex M0 core, 128 kB of SRAM, optical boot loader, and a variety of additional circuitry to interface with sensors.  All clocks and RF signals on \scum{} are generated using free-running on-chip oscillators.  Traditional wireless systems utilize an off-chip crystal oscillator reference, calibrating the on-chip oscillators from the crystal using a Phase-Locked Loop (PLL).  External crystal oscillators are antithetical to the design goals of \scum{}, as they increase device cost and power consumption.  As a result, the oscillators used by \scum{} are less precise and reliable than those of traditional devices, although they are fully compatible with existing wireless protocols with the use of calibration techniques \cite{burnett_free-running_2021, moreno_frequency_2022, chang_surviving_2022}.  Crystal-free devices operating in the 2.4~GHz ISM band using BLE-like wireless protocols \cite{alghaihab_307_2020, zhao_24-ghz_2021, shen_crystal-less_2025} have been published, though to our knowledge, a crystal-free BLE transceiver has not yet been demonstrated.

The development of crystal-free devices like \scum{} has been difficult due to the lack of existing design tools with low-level access, making pre-silicon validation challenging and increasing the likelihood of flawed or failing devices designed using these architectures. 
These are unacceptable risks when cost is the primary concern for these devices.

With the motivation of creating a development platform suitable for crystal-free device design in mind, our goals for this system are as follows.  We are only aiming for BLE revision 4.0 PHY- and MAC-layer designs. Newer BLE revisions require an initial connection to be made on the 4.0 PHY, and higher layers of the stack are less dependent on the particular device architecture and are predominantly programming exercises rather than device design. Open-source BLE stacks such as BlueKitchen \cite{noauthor_bluekitchen_nodate} are available to implement the rest of the BLE system. An open-source reference platform utilizing commercially available devices is necessary to allow new collaborators interested in starting crystal-free research to get started. Such an approach is also viable for others interested in developing similar platforms for other architectures. Our platform also aims to be inter-compatible with existing \scum{} devices and to be as modular as possible. This enables partially working \scum{} designs to be useful in future research, as well as more direct comparisons of crystal-free architectures with traditional architectures.

The rest of this paper is organized as follows. Section \ref{Design} details the relevant BLE 4.0 specifications and our overall platform design. Section \ref{Implementation} discusses simulations of the protocol and design architecture, as well as the particular designs created as a demonstration of this platform. Section \ref{Results} contains a demonstration of BLE rev. 4.0 functionality and details some important receiver specifications, including comparisons of our design running on hardware both with and without a crystal oscillator reference. Finally, Section \ref{Conclusion} further discusses the importance of this work and details some of our intended future uses of this platform.

%% file: design.tex
Co-design of the digital twin platform and hardware architecture is critical. Many of the architecture choices described here are a direct consequence of the goal of using \scum{} as a hardware front end to achieve the most realistic noise environment possible. 

\subsection{Why is Crystal-Free Difficult?} \label{Crystal Free}
\begin{figure}
    \centering
    \includegraphics[width=.49\textwidth]{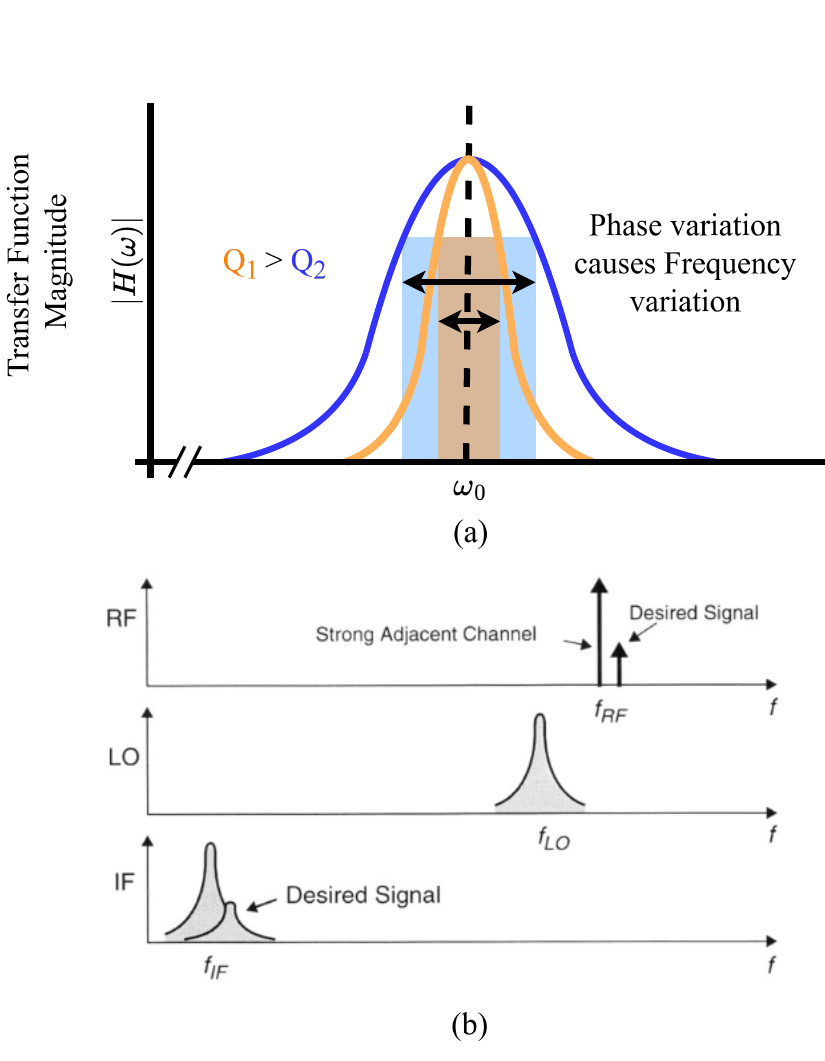}
    \caption{(a) Phase noise comparison of two oscillators with different quality factors, Q1 (e.g, off-chip crystal reference) and Q2 (e.g, on-chip oscillator) \cite{rohail_tutorial_2024} (b) Effect of increased phase noise on communication system performance \cite{hajimiri_design_1999}}
    \label{fig:phase-noise}
\end{figure}

\scum{}'s crystal-free architecture and existing hardware capabilities place fundamental limitations on this system. However, these constraints have guided this work to follow the crystal-free ethos of minimizing hardware in pursuit of power and cost efficiency. Traditional architectures utilize off-chip crystal oscillators and Phase-Locked Loops to improve the phase noise performance of the on-chip oscillators used in RF signal processing. As shown in Figure \ref{fig:phase-noise}, the phase noise profile as a result of eliminating these components leads to frequency uncertainty and is of particular concern in FSK-type systems such as BLE. As a result, \scum{} uses a Low-IF architecture. This shifts the intended signals out of the local oscillator's (LO) flicker noise region and integrates less of the LO phase noise when down-converting. While this approach improves transceiver performance for crystal-free systems, their performance still falls short of traditional systems with crystal oscillator references. However, as shown in \cite{hippe_digital_2024}, there is only a limited set of parameters that have a significant impact on receiver performance, and careful system design can allow these systems to exceed the necessary performance for many common wireless communication protocols. As will be shown in Section \ref{Results}, the unique design space as a result of these choices allows for minimal hardware implementations to achieve receiver sensitivity results up to \bestsens{~dBm}.

\subsection{Relevant Bluetooth Low Energy Specifications}
Bluetooth revisions 4.0 and newer, often referred to as Bluetooth Low Energy (BLE), are a set of specifications for low data rate, short-range communications \cite{noauthor_core_2024}. These specifications provide a range of data rates, error correction codes, and packet lengths, among other things. Any 4.0 or newer protocol, however, makes an initial connection using the 4.0 specification. Therefore, the 4.0 specification is the focus of this work. The techniques described here are directly applicable to designs for other specifications.

The BLE 4.0 specification defines the Bluetooth LE1M PHY, allowing devices to communicate as 1~Mb/s (peak) on the 2.4~GHz ISM band. Modulation is performed using Gaussian Frequency Shift Keying (GFSK), a form of Minimum Shift Keying (MSK) that minimizes out-of-band power. In the BLE 1M PHY, the carrier frequency is modulated by $\pm$250~kHz to transmit digital data, with bit transitions smoothed using a Gaussian filter. This modulation scheme provides some unique properties when implemented in non-standard ways, shown in this and previous work, informing the receiver architecture detailed in Section \ref{PHY}.

The BLE 4.0 spec also defines 40 channels in the 2.4~GHz ISM band, beginning at 2.402~GHz and ending at 2.480~GHz, with a channel spacing of 2~MHz. 37 of these channels (0-36) are "Data" channels used after a connection has been made and are spaced between the three "Advertising" channels (37 [2.402~GHz], 38 [2.426~GHz], and 39 [2.480~GHz]). As the name suggests, peripheral devices advertise on these channels for host devices to initiate connections. This process is further described in Section \ref{Link Layer}.

\subsection{PHY-Layer Design}\label{PHY}

\begin{table}[]
\begin{threeparttable}[]
\begin{tabular}{|c|c|c|}
\hline
\textbf{PHY} & \begin{tabular}[c]{@{}c@{}}\textbf{IEEE 802.15.4}\\ \textbf{2.4 GHz}\end{tabular} & \begin{tabular}[c]{@{}c@{}}\textbf{Bluetooth}\\ \textbf{LE1M}\end{tabular} \\ \hline
Band & \begin{tabular}[c]{@{}c@{}}2.405 GHz to\\ 2.480 GHz\end{tabular} & \begin{tabular}[c]{@{}c@{}}2.402 GHz to\\ 2.480 GHz\end{tabular} \\ \hline
\begin{tabular}[c]{@{}c@{}}Channel\\ Spacing\end{tabular} & 5 MHz & 2 MHz \\ \hline
\begin{tabular}[c]{@{}c@{}}Bandwidth\\ (peak)\end{tabular} & 250 Kbps & 1 Mbps \\ \hline
\begin{tabular}[c]{@{}c@{}}Modulation \\ Type\end{tabular} & \begin{tabular}[c]{@{}c@{}} Half-Sine Shaping \\ Offset-Quadrature \\ Phase Shift Keying\\ (HSS O-QPSK) \tnote{1}\end{tabular} & \begin{tabular}[c]{@{}c@{}}Gaussian Frequency\\ Shift Keying\\ (GFSK)\end{tabular} \\ \hline
\begin{tabular}[c]{@{}c@{}}Tone Spacing\\ (Nominal)\end{tabular} & 1 MHz ($\pm$ 500 kHz) & 500 kHz ($\pm$ 250 kHz) \\ \hline
\begin{tabular}[c]{@{}c@{}}Symbol/Chip \\ Rate\end{tabular} & 2 MChips/s & 1 MSps \\ \hline
\begin{tabular}[c]{@{}c@{}}Bits per \\ Symbol/Chip\end{tabular} & \begin{tabular}[c]{@{}c@{}}0.125 bits per Chip\\ (4 bits = 32 chips)\end{tabular} & 1 bit per Symbol \\ \hline
\begin{tabular}[c]{@{}c@{}}Timing \\ Indicator\end{tabular} & \begin{tabular}[c]{@{}c@{}}40-bit Sync. Header\\ (160~\textmu s)\end{tabular} & \begin{tabular}[c]{@{}c@{}}8-bit Preamble\\ (8~\textmu s)\end{tabular} \\ \hline
\begin{tabular}[c]{@{}c@{}}Maximum RX \\ Error Rate\end{tabular} & \begin{tabular}[c]{@{}c@{}}6.5\% Chip Error\\ (1\% Packet Error)\end{tabular} & \begin{tabular}[c]{@{}c@{}}0.1\% Bit Error\\ (30.8\% Packet Error)\end{tabular} \\ \hline
\end{tabular}
\begin{tablenotes}
    \item[1] HSS O-QPSK is equivalent to MSK \cite{Notor2003CMOSRA}
\end{tablenotes}
\caption{Comparison of IEEE 802.15.4 2.4 GHz PHY and BLE 1M PHY. \scum{} has been demonstrated previously as a fully capable 802.15.4 transceiver.}
\label{tab:phy}
\end{threeparttable}
\end{table}

\scum{}'s RF front end was designed for IEEE 802.15.4 \cite{noauthor_ieee_2020}, and is capable of communicating with COTS 802.15.4-compliant devices. At the physical layer, IEEE 802.15.4 and BLE are similar enough to warrant comparison. These standards are compared in Table \ref{tab:phy}.

In Burnett et al. \cite{burnett_free-running_2021}, \scum{} is shown to have a receiver sensitivity of -83~dBm for 802.15.4 communications. Although this does not meet the -85~dBm specification of IEEE 802.15.4, \scum{} is practically able to perform data exchange with commercially available 802.15.4-capable devices. The similarity of the IEEE 802.15.4 2.4~GHz and BLE 1M physical layers indicates that similar capabilities should be possible for BLE. BLE's slower symbol rate and Gaussian filtering should improve performance in comparison to 802.15.4, though the lack of error correction and much shorter timing indicator could cause difficulties. 

Among the unique features of the proposed system is bit-level control over the BLE PHY. The most common BLE implementations used for the development of IoT devices do not allow access below the Host-Controller Interface (HCI). However, this access is necessary for designing novel wireless architectures. This section details the relevant hardware specifications of \scum{} and the ADALM-PLUTO SDR, along with the signal processing algorithms used to transmit and receive BLE packets.

\subsubsection{Transmitter Implementation}
This system is predominantly focused on the design of crystal-free BLE receivers, as BLE transmitter capability has already been demonstrated \cite{yuan_temperature-compensated_2022}. However, transmitting is a necessary part of this system to demonstrate full transceiver capability. Three transmitter implementations have been created for use with this platform. The first exists above the PHY layer and is implemented using a Digilent Analog Discovery 2 \cite{noauthor_analog_nodate} interfaced with an Agilent E4438C Vector Signal Generator. This implementation uses the data and data clock inputs of the Vector Signal Generator to take digital data provided by the Analog Discovery 2 (e.g., a packet) and modulates the data using the programmed BLE modulation scheme. The final two implementations, on the ADALM-PLUTO and \scum{}, respectively, modulate using in-phase and quadrature (I/Q) samples generated by the transmitter script. These implementations are discussed in more detail in Section \ref{Implementation}.

\subsubsection{Receiver Implementation}
As discussed in Section \ref{Crystal Free}, \scum{} uses a non-coherent Low-IF wireless architecture. While transmission with these types of architectures is relatively simple, receiving signals requires more careful analysis of the design space \cite{hippe_digital_2024}. \scum{}'s receiver front end contains 4-bit In-Phase and Quadrature ADCs sampling the down-converted IF signal (nominally 2.5~MHz in 802.15.4 operation) at 16~MHz. I/Q samples from the baseband ADCs are provided to the demodulation and clock recovery algorithms detailed here.

This work utilizes the Low-IF GFSK design space analysis work detailed in \cite{hippe_digital_2024}. The matched filter algorithm presented was implemented in Python, MATLAB, C, and SystemVerilog RTL, with each output compared to ensure correctness. These implementations are parameterizable for the various hardware platforms used in this system. The clock recovery algorithm used in this work is a bit-transition detector, as the matched filter implementation was found to have transition times within a few symbols, even in the presence of noise. An I/Q sample-based implementation based on \cite{dandrea_digital_1990}, with modifications made for the Low-IF used, was also implemented but was not found to match the performance of the bit-transition based clock recovery algorithm. We suspect this is due to this algorithm's slow convergence relative to the preamble of a BLE packet. The bit-transition-based algorithm does not require any multiplication operations, which are needed in the I/Q sample-based approach. This difference should simplify hardware implementations. Simulations of this receiver algorithm with Gaussian amplitude noise to verify their robustness, though these results are not shown here.

\subsection{MAC/Link Layer Design}\label{Link Layer}
\subsubsection{Advertising Events}
The initiation of any BLE communications begins with peripheral devices advertising their existence. There are a variety of advertising packet types devices can use, though the focus of this work is on Indirected Advertising packets (ADV\_IND). These packets indicate that a peripheral device is connectable from a host device that hears its advertisement. There are two common ways host devices can scan for these packets: Passive Scanning or Active Scanning. The Bluetooth specification does not specify that devices have to use one or both of these schemes, and it was found during testing that devices from different manufacturers differ in their requirements for displaying a device as connectable. However, the packet transmission events that occur during passive scanning are a subset of those used in active scanning, so any device that can participate in an active scan event can participate in passive scanning as well. Active scanning events are also the simplest BLE events that require transceiver functionality and, as such, are the focus of the demonstrations in this work.

\begin{figure}[h]
    \centering
    \includegraphics[width=0.49\textwidth]{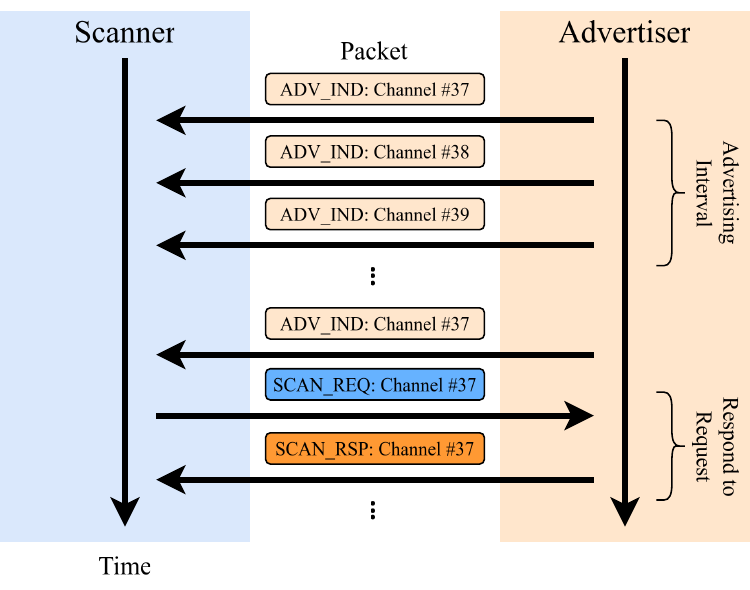}
    \caption{Event diagram for active scanning. Though not specified in the BLE specification, most commercially available devices use this event sequence to search for connectable BLE devices.}\label{fig:active-scanning}
\end{figure}

An event diagram for active scanning is shown in Figure \ref{fig:active-scanning}. This sequence can occur on any of the advertising channels, and the peripheral may advertise on any or all of these channels in turn. When a host device is actively scanning, it will respond to any indirected advertising packets it receives with a scan request packet (SCAN\_REQ), which is directed to the peripheral device using its Advertising Address. This packet serves to request more information from the peripheral device. When the peripheral device receives a scan request packet addressed to it, it should respond in time with a scan response packet (SCAN\_RSP). This packet does not need to contain any additional data than what is included in the initial packet, though we have included additional data to confirm the transaction has occurred. If the transaction fails, the host may filter out the peripheral and not display it as connectable to the user.

\subsubsection{Bluetooth Rev. 4.0 Packet Structure}

The general Bluetooth Revision 4.0 packet structure and specifics related to the types used in later demonstrations are worth detailing here, as they will be relevant towards testing performed in Section \ref{Results}. Standard BLE packets contain three fields: Header, Protocol Data Unit (PDU), and Cyclic Redundancy Check (CRC). Each of these fields is discussed in further detail as follows.

The BLE packet header consists of two sub-fields: an 8-bit preamble and a 32-bit access address. The preamble is a sequence of alternating bits (0x55 or 0xAA), defined such that the last transmitted bit of the preamble differs from the first transmitted bit of the access address. This guarantees that each packet begins with a sequence of eight consecutive bit transitions and is predominantly used for clock synchronization in the receiver. The preamble conveys no data and is not required to be correctly received and is not included as part of the packet for the packet error rate testing performed in Section \ref{Results}. The access address is transmitted least-significant-bit first and is always the same for advertising packets (0x8E89BED6). 

Data is included in specific blocks in the PDU. The PDU begins with a 2-byte header field, which indicates the type of packet, device flags, and the packet payload length in bytes. These fields are shown in Table \ref{tab:pdu} in transmission order, with each field transmitted least significant bit first. 

\begin{table*}[t]
    \centering
    \begin{tabular}{lclllclccclllllllllclclcll}
    \cline{1-17} \cline{19-26}
    \multicolumn{1}{|l|}{\textbf{Bit Index}} & 0                    & \multicolumn{1}{c}{} & \multicolumn{1}{c}{} & \multicolumn{1}{c|}{3} & 4                    & \multicolumn{1}{c|}{5} & \multicolumn{1}{c|}{6}     & \multicolumn{1}{c|}{7}     & 8                    & \multicolumn{1}{c}{} & \multicolumn{1}{c}{} & \multicolumn{1}{c}{} & \multicolumn{1}{c}{} & \multicolumn{1}{c}{} & \multicolumn{1}{c}{} & \multicolumn{1}{c|}{15} & \multicolumn{1}{l|}{} & \multicolumn{1}{l|}{\textbf{Bit Index}} & 0                    & \multicolumn{1}{c|}{7} & 8                    & \multicolumn{1}{c|}{15} & 16                   & \multicolumn{1}{c}{...} & \multicolumn{1}{c|}{$\le$247} \\ \cline{1-17} \cline{19-26} 
    \multicolumn{1}{|l|}{\textbf{Field}}     & \multicolumn{4}{c|}{PDU Type}                                                               & \multicolumn{2}{c|}{RFU}                      & \multicolumn{1}{c|}{TxAdd} & \multicolumn{1}{c|}{RxAdd} & \multicolumn{8}{c|}{Length}                                                                                                                                                              & \multicolumn{1}{l|}{} & \multicolumn{1}{l|}{\textbf{Field}}     & \multicolumn{2}{c|}{Length}                   & \multicolumn{2}{c|}{GAP Code}                  & \multicolumn{3}{c|}{Data}                                                      \\ \cline{1-17} \cline{19-26} 
                                    & \multicolumn{1}{l}{} &                      &                      &                        & \multicolumn{1}{l}{} &                        & \multicolumn{1}{l}{}       & \multicolumn{1}{l}{}       & \multicolumn{1}{l}{} &                      &                      &                      &                      &                      &                      &                         &                       &                                & \multicolumn{1}{l}{} &                        & \multicolumn{1}{l}{} &                         & \multicolumn{1}{l}{} &                         &                               \\
    \multicolumn{17}{c}{(a)}                                                                                                                                                                                                                                                                                                                                                                                                           &                       & \multicolumn{8}{c}{(b)}                                                                                                                                                                                         
    \end{tabular}
    \caption{(a) PDU Header Fields. (b) GAP Block Fields.}
    \label{tab:pdu}
\end{table*}

The payload consists of a minimum of 6 and a maximum of 37 bytes of data. The first 6 bytes of this are the device's access address, transmitted least significant bit first. The remaining bytes may be filled with any number of payload blocks (Generic Access Profile [GAP] or Generic Attribute Profile [GATT]). GAP blocks are the only types relevant to advertisement packets and are discussed here. Each GAP block is structured as shown in Figure \ref{tab:pdu}. The length and GAP code are 1 byte each. The length value is the length of the block, excluding the length byte. GAP blocks are transmitted least significant bit first per byte, with the order of bytes maintained.

BLE 4.0 utilizes a 24-bit CRC with polynomial 0x00065B initialized to 0x555555 and is calculated for the PDU only in transmission order. After the CRC is computed, the PDU and CRC are "whitened" by xor-ing with the most significant bit of a 7-bit LFSR with polynomial 0x11 and initialized using the channel number. The channel number is placed, most significant bit first, in positions 1-6, with bit 0 always initialized to 1. Applying the same process at the receiver "dewhitens" the packet. This process ensures that the transmitted data has approximately evenly distributed bit transitions to help the receiver perform clock recovery.

The maximum BLE 4.0 packet is 376 bits long (368 after excluding the preamble). This is the default packet length referred to for any packet-error-rate to bit-error-rate conversions (or vice versa) in Section \ref{Results}.

%% file: implementation.tex
Two important benefits of using a digital twin in developing wireless devices are rapid iteration and testing with realistic noise profiles. However, in many applications, especially crystal-free architectures, modeling a wireless device and channel can be highly complex, computationally intensive, or infeasible, even with modern simulation models \cite{rohail_tutorial_2024}. As such, our proposal is a "hybrid" digital twin, where certain subsystems are physically implemented rather than simulated in great detail, though simulation representations are still used in our proposed ecosystem. To address both rapid iteration and realistic noise profiles, this ecosystem was developed with two hardware platforms in mind: a COTS reference platform implemented in an Analog Devices ADALM-PLUTO SDR \cite{noauthor_adalm-pluto_nodate} and a \scum{} RF front end interfaced with the digital baseband in an FPGA.

\subsection{Commercial Reference Platform}

Developing a commercial reference platform enables inexpensive and rapid iteration of more complex implementations using \scum{}. The ADALM-PLUTO features a high-quality dual-channel transceiver that utilizes 12-bit ADCs and DACs at up to 61.44 MSPS, exceeding the current hardware capabilities of \scum{}. The transceiver is paired with a Zynq 7000 series SoC from Xilinx \cite{noauthor_amd_nodate}, allowing for embedded firmware and custom RTL to be developed and tested.

Modulation and demodulation scripts were written using Python to interface with ADALM-PLUTO using the PySDR library \cite{noauthor_pysdr_nodate} to test the feasibility of our implementations quickly. However, performing demodulation in Python was insufficient to demonstrate a connectable BLE transceiver and perform receiver sensitivity testing in a realistic time frame.

To enable real-time continuous decoding processing on the ADALM-PLUTO, each decoder function was rewritten in Embedded C. Leveraging Analog Devices’ libiio library \cite{noauthor_libiio_nodate}, these functions were integrated into the ADALM-PLUTO platform. The implemented demodulation algorithm was tested using two different methods. In Method 1, an Analog Discovery 2 \cite{noauthor_analog_nodate} and an E4438C Vector Signal Generator (VSG) were used to modulate and transmit BLE packets. In Method 2, another ADALM-PLUTO was used as a simple, low-cost transmitter. This setup provides a flexible testing platform and enables control over PHY-layer parameters for transmitting, such as frequency, symbol rate, packet content, repeat counts, and intervals. These configurations are programatically set, eliminating the need for manual adjustments. 

During the development of the commercial reference platform, we ran into constraints that affected our receiver performance and forced changes to the implementation. At first, we tried to run the entire demodulation algorithm in firmware on the ADALM-PLUTO’s Cortex-A9 processor, but it was found to be infeasible. We also encountered issues with USB communications between the ADALM-PLUTO and the host PC. Due to USB 2.0 speed limits and how the Libiio API handles its receive buffer, continuous receiving of samples was not possible without interruption, and therefore, data loss. The libiio API only refills the hardware’s receive buffer after all data in the previous buffer has been transferred over USB. This introduces a delay between buffer refills, causing some samples to be missed during those gaps. Fortunately, our real-world tests showed that each buffer contains about 125~ms of data with a gap of approximately 5~ms between buffer refills. As such, we can capture most of the I/Q samples, though this does place a limit on our receiver's sensitivity. To reduce timing issues and avoid extra delays between receptions, the USB communication and the demodulation algorithm are run in separate threads. Furthermore, when the ADALM-PLUTO was used as the transmitter, turning off the ADALM-PLUTO's transmitter between transmissions caused issues, leading to missing packets. Instead, the ADALM-PLUTO transmits continuously between packets without modulating data. After completing sensitivity testing, we combined our transmitter and receiver into transceiver firmware to run on the ADALM-PLUTO to demonstrate active scanning events. These results are shown in Section \ref{Results}.

\subsection{\scum{} \& Digital Baseband in FPGA}
The design of the BLE digital baseband was inspired by the IEEE 802.15.4 implementation on the existing \scum{} hardware. Simple changes were made to convert the implementation to BLE. The matched filter needed a new template to match the shift in frequency deviation. Additionally, the buffer required an increase to accommodate the 1~$\mu$s bit time for BLE compared to the 0.5~$\mu$s chip time for IEEE 802.15.4. The digital baseband was translated into SystemVerilog RTL to be tested and synthesized on the ADALM-PLUTO's FPGA and contributed to an upcoming tapeout. The developed RTL is essential for use with existing \scum{} devices. \scum{}'s ADC output values can be routed directly to GPIOs, allowing for the existing IEEE 802.15.4 digital baseband to be bypassed in favor of a new digital baseband implemented in an FPGA. This is a crucial design feature for any novel RFIC, and when used in this manner with a digital twin, it allows for the most realistic noise profiling for novel RFIC architectures. The ability to quickly swap from the ADALM-PLUTO as the RF front end to \scum{} allows us for the first time to compare crystal oscillator and crystal-free oscillator performance as applied to sensitivity measurements, as detailed in Section \ref{Results}. \scum{} has a more constrained analog front-end architecture than the ADALM-PLUTO, so extrapolation of the feasibility of crystal-free architectures in BLE transceivers demonstrated in this work to higher ADC bit depths or sampling frequencies are among potential next directions of this work. Although not tested for this work, the ADALM-PLUTO does feature an external clock pin, which could be used to fully test the effects of higher sampling frequencies and ADC bit depths using noisy oscillators.

\subsubsection{\scum{} RX Simulation}
To verify that the external digital baseband implementation was correct, IEEE 802.15.4 decoding was built and tested on the FPGA and simulated in MATLAB. 
The \scum{} chip was designed for this wireless communication standard and the preliminary step should demonstrate no change in bit error performance when moving the signal processing off-chip.
The MATLAB digital baseband code was designed to mirror the hardware implementation of the matched filter and timing recovery. The test configuration is represented in Figure \ref{fig:testing}b. 
The simulation data included I/Q samples produced from the \scum{} RF front end. 
IEEE 802.15.4 packets were transmitted using PLUTO, and the I/Q samples were recorded using a Digital Discovery \cite{noauthor_digital_nodate}. 
After being correctly decoded, the same samples were used in Vivado simulations to confirm the functionality of the FPGA hardware design.
This process was repeated for the BLE version of the implementation.

%% file: results.tex
\begin{figure}[]
    \centering
    \includegraphics[width=0.49\textwidth]{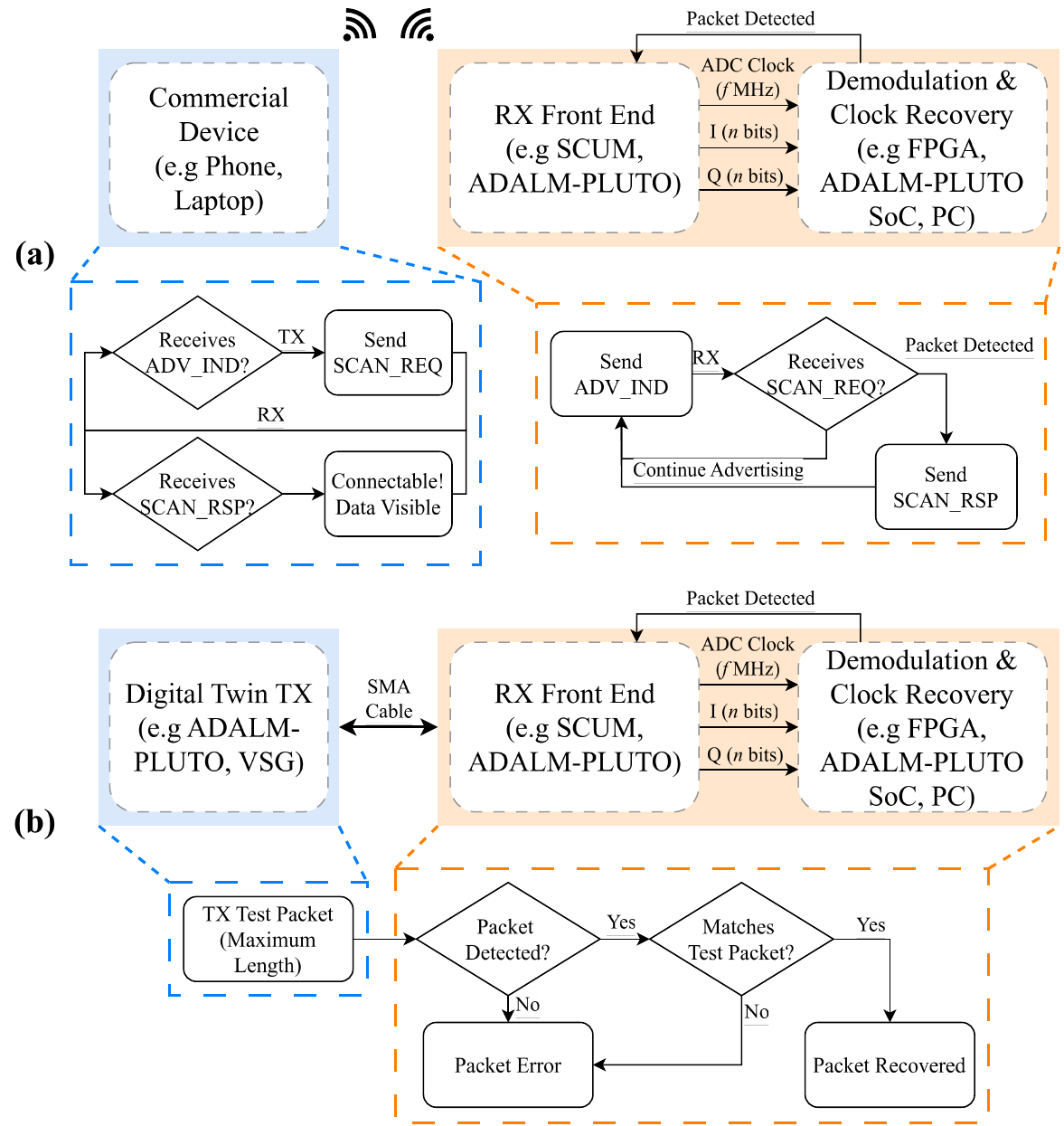}
    \caption{ (a) Block and flow diagrams of the connectable demo. The proposed platform initiates the process by transmitting an advertisement packet. When a scan request packet is sent from the phone, the platform switches to sending a scan response packet with additional data. (b) Block and flow diagrams of the performed receiver sensitivity testing. One of the transmitter implementations transmits a maximum-length packet. If the packet is detected and matches the expected packet, the packet is counted as correctly recovered. Otherwise, the packet is counted as an error. This packet error rate is later converted to the equivalent bit error rate.}
    \label{fig:testing}
\end{figure}

Crystal-free wireless architectures make receiver phase noise trade-offs in pursuit of lower power consumption and device cost. Such trade-offs have a direct impact on the receiver sensitivity of the system. However, we can demonstrate the first crystal-free BLE transceiver system as a result of the design enabled by the proposed digital twin architecture as detailed in Section \ref{Design}. This is accomplished through two approaches: a qualitative approach, demonstrating an active scanning event, and a quantitative approach through receiver sensitivity testing.

\subsection{Demonstration as a Connectable BLE Device}
\begin{figure*}[tb]
    \centering
    \includegraphics[width=0.99\textwidth]{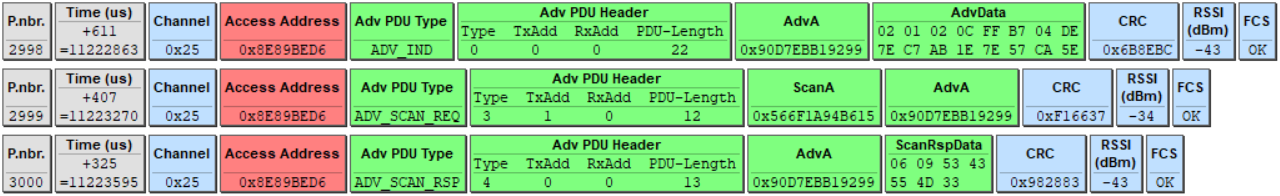}
    \caption{Packet sniffer \cite{noauthor_cc2540_nodate} showing packets of active scanning event performed by a commercially available device and the proposed platform, in this case, the ADALM-PLUTO SDR.}
    \label{fig:packet_sniff}
\end{figure*}

Active scanning is detailed in Figure \ref{fig:active-scanning}. Active scanning is the simplest event that requires both host and peripheral devices to have transceiver capabilities and is verifiable using a BLE scanner app on a host device \cite{noauthor_simplicity_nodate, noauthor_nrf_nodate} or packet sniffer \cite{noauthor_cc2540_nodate}. This approach verifies that crystal-free wireless systems such as \scum{} are "standards compatible" and capable of communicating with commercially available BLE devices. Figure \ref{fig:connectable} shows \scum{} identifying as a connectable device, and Figure \ref{fig:packet_sniff} shows the packets transmitted by both devices during an active scanning event, as detected by a packet sniffer.

The live BLE \scum{} demonstration is the device that initiates the connection using an advertisement packet. Once the phone has received the advertisement packet, it will send a scan request packet. Currently, the chain ends with \scum{} decoding the scan request packet. However, the next step in the process would be for \scum{} to send the scan response packet. If the phone correctly receives this packet, the data within this packet should be visible from the mobile device. A packet sniffer is used to confirm that the packet received by \scum{} is the one being send by the phone. The scan address is the main field in the scan request packet that is used to identify the device. 

As a result of this demonstration, \scum{} is shown capable of bi-directional communications with commercially available BLE devices. Future \scum{} revisions with BLE hardware will require further development in the link layer and higher layers of the BLE stack, potentially using open-source stack software such as BlueKitchen \cite{noauthor_bluekitchen_nodate}.

\subsection{Receiver Sensitivity \& Comparison}

\begin{figure}[]
    \centering
    \includegraphics[width=0.49\textwidth]{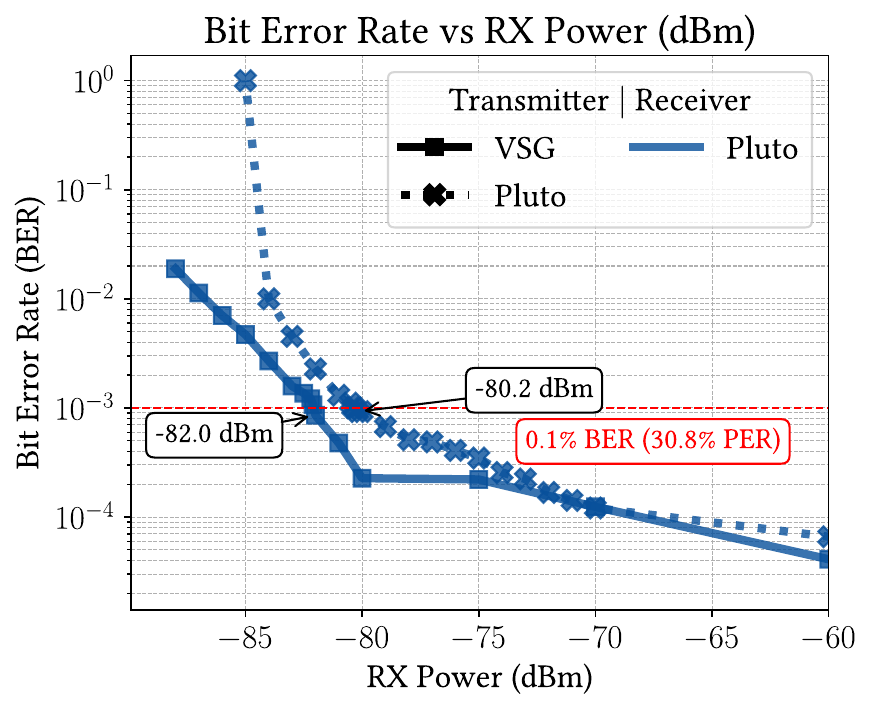}
    \caption{Received power vs bit error rate (BER) for the proposed digital twin system. All configurations exceed the minimum Bluetooth 4.0 receiver sensitivity specification (-70~dBm).}\label{fig:waterfall}
\end{figure}

Receiver sensitivity is the predominant factor in the effective range of the system and, therefore, is among the most important specifications for communication quality. The Bluetooth LE1M PHY defines receiver sensitivity as the minimum received power at which the bit error rate (BER) is less than 0.1\% \cite{noauthor_core_2024}. However, for testing purposes, packet error rates (PER) are used, with approximately 30.8\% PER equivalent to 0.1\% BER. This is measured over at least 1500 maximum-length packets (368 bits, after excluding the preamble). We performed these tests using 2000 maximum-length packets for both possible RF front ends as described in Figure \ref{fig:block_diagram}. The results of these tests are shown in Figure \ref{fig:waterfall}.

\begin{table*}
\begin{threeparttable}[]
\begin{tabular}{|l|l|l|l|l|l|l|}
\hline
                                                                    & \textbf{\begin{tabular}[c]{@{}l@{}}\scum{} \cite{maksimovic_crystal-free_2019} +\\ This work\end{tabular}}     & \begin{tabular}[c]{@{}l@{}} MWTL \\ 2025 \cite{wang_high-robust_2025} \end{tabular} & \begin{tabular}[c]{@{}l@{}} CICC \\ 2017 \cite{shirazi_980w_2017} \end{tabular}      & \begin{tabular}[c]{@{}l@{}} T-MTT \\ 2019 \cite{silva-pereira_17-mw_2019} \end{tabular}      & \begin{tabular}[c]{@{}l@{}} ISSCC \\ 2024 \cite{scolari_232_2024} \end{tabular}                      & \begin{tabular}[c]{@{}l@{}} A-SSCC \\ 2016 \cite{oshiro_32_2016} \end{tabular}     \\ \hline
Technology                                                          & \textbf{65~nm CMOS}                                                & 65~nm CMOS                                                   & 40~nm CMOS                                              & 130~nm CMOS & 22~nm FDSOI                                                & 65~nm CMOS \\ \hline
Comm. Type                                                          & \textbf{\begin{tabular}[c]{@{}l@{}}802.15.4 TRX,\\ BLE TX\end{tabular}} & BLE RX                                                            & BLE RX                                                       & BLE RX           & \begin{tabular}[c]{@{}l@{}}802.15.4 TRX,\\ BLE TRX\end{tabular} & BLE TRX         \\ \hline
Architecture     & \textbf{\begin{tabular}[c]{@{}l@{}}Crystal-Free \\ Low-IF\end{tabular}} & Low-IF                                                            & \begin{tabular}[c]{@{}l@{}}Direct \\ Conversion\end{tabular} & Low-IF           & SDR                                                & Sliding-IF      \\ \hline
\begin{tabular}[c]{@{}l@{}}RX P$_{dc}$\\ {[}mW{]}\end{tabular}                                                    & \textbf{1.03} \tnote{1}                                                          & 6                                                                 & 0.98                                                         & 1.7              & 3.31                                                            & 9.6             \\ \hline
\begin{tabular}[c]{@{}l@{}}RX Area\\ {[}mm$^2${]}\end{tabular}                                                       & \textbf{1.48} \tnote{2}                                                          & 1.2                                                               & 0.7                                                          & 0.7              & 1.0                                                               & 2.63 \tnote{3}            \\ \hline
\begin{tabular}[c]{@{}l@{}}RX Sensitivity \\ {[}dBm{]}\end{tabular} & \textbf{-83~dBm} \tnote{4}                                                               & -96                                                               & -95.8                                                        & -92              & -98.2                                                           & -93             \\ \hline
\end{tabular}
\begin{tablenotes}
    \item[1] 802.15.4 analog/RF power only. Others include digital baseband power.
    \item[2] RF Front End and 802.15.4 Digital Baseband.
    \item[3] RF Front End only.
    \item[4] 802.15.4 RX sensitivity. We postulate \scum{} should have similar RX sensitivity for BLE, as shown from the PHY comparison in Table \ref{tab:phy}.
\end{tablenotes}
\caption{Comparison of the proposed BLE Digital Twin using \scum{} as the RF Front End and Digital Baseband in an FPGA to published BLE Receivers.}
\label{tab:comparison}
\end{threeparttable}
\end{table*}

We show our proposed platform achieves a receiver sensitivity of \plutosens{}{~dBm}, exceeding the minimum BLE receiver sensitivity of -70~dBm. We estimate \scum{} should obtain similar performance due to \scum{}'s -83~dBm receiver sensitivity for IEEE 802.15.4 and to the similarity between the IEEE 802.15.4 2.4~GHz PHY and the Bluetooth LE1M PHY. We estimate of \scum{}'s BLE performance metrics based on its 802.15.4 metrics and compare with state-of-the-art BLE receivers in Table \ref{tab:comparison}. While the obtained sensitivity metric does not meet the performance of published literature, it meets the minimum BLE receiver sensitivity metric. Furthermore, \scum{} is the only architecture compared that is capable of monolithic integration, allowing \scum{} to be more capable of operating in extreme environments (e.g. high/low temperature, high magnetic field, high radiation) than other systems.

%% file: conclusion.tex
In this work, we propose a low-level hybrid digital twin reference platform for the design of crystal-free BLE devices. The proposed system is focused on the PHY and MAC levels of the BLE stack and features an open-source implementation, enabling new device designs utilizing novel wireless communication architectures, such as the crystal-free architecture featured in this work, to be developed at a lower cost and with better validation than before. 

As demonstrated in this work, the proposed platform is useful in a wide exploration of the design space. This has allowed us to demonstrate an adaptable BLE transceiver that can be optimized for a variety of different hardware implementations.  Furthermore, the benefits of this approach can contribute to reducing the cost of the final design, which is of the utmost importance for IoT devices intended for environmental sensing applications. 

However, this platform still has a few limitations. The particular hardware used for demonstration in this work is limited in the interest of maintaining a low cost, and further work would be required for those looking to utilize this platform for the design of higher-performance BLE transceivers. Further efforts to simplify the use of this platform would be beneficial for broader applications than those detailed in this work.

We have used the proposed system to demonstrate the first, to our knowledge, crystal-free Bluetooth Low Energy transceiver capable of bi-directional communications with commercially available devices. This capability is shown through the transceiver's participation in active advertising events. As such, this transceiver is "standards compatible" and capable of data exchange with commercially available BLE devices. We envision future \scum{} designs to be equipped with our BLE transceiver design, improving the usability of \scum{} as a widely adaptable environmental sensing platform. We have also compared the performance of our receiver using the ADALM-PLUTO with Bluetooth specifications and other published works, finding its \plutosens{~dBm} receive sensitivity to be better than the minimum specification, though worse than other published receivers. We believe \scum{} should attain a similar receiver sensitivity, though this testing is still to be performed.

We aim to apply the proposed platform and transceiver detailed in this work to explore crystal-free BLE further. This could include a demonstration of sensor data communication with a custom phone or laptop application. A demonstration like this would enable \scum{} to have a smoother initial deployment experience: connect a sensor, write interfacing code with the sensor, and send the collected data using the available BLE hardware/software. This experience would closely match that of hobbyist-type devices and enable the design of higher-level sensor systems using \scum{}. Furthermore, integration with a Bluetooth stack, such as BlueKitchen \cite{noauthor_bluekitchen_nodate}, would provide traditional pairing and communication exchange with commercial devices. 

Lastly, we also envision more out-of-the-box experimentation with BLE as a communication protocol for environmental sensing applications. For example, it would be more efficient for a low-power device such as a sensor mote to act as a host device, listening for available devices to transmit its data to, rather than acting as a peripheral. As a peripheral device, any advertisement that does not lead to a connection or data transmission is a waste of energy and limits the device's lifespan. We imagine a device intending to connect to \scum{} including some special data in its advertising packet, which, when received, awakens \scum{} to initialize a proper connection and transmit sensor data. We see this as important in streamlining the process of deployment and use of wireless sensor networks in many environmental sensing applications.